\documentstyle{article}

\begin{document}
% \draft
% \preprint{ }
\title{A Micro-Thermodynamic Formalism}
\author{Hans Henrik Rugh \\
        Department of Mathematics,
          University of Cergy-Pontoise, \\
	  France 95302}
\date{\today}
\maketitle

\begin{abstract}
We consider the micro-canonical
ensemble of a classical Hamiltonian dynamical system,
the Hamiltonian being parameter dependent and in the possible presence
of other first integrals.
We describe a thermodynamic formalism in which
a 1st law of thermodynamics, or fundamental relation,
is based upon the bulk-entropy,
$S_\Omega$. Under an ergodic hypothesis,
$S_\Omega$ is shown to be  an adiabatic invariant.
Expressions for derivatives  and
thermodynamic relations are derived
within the micro-canonical ensemble itself.
\end{abstract}
%\pacs{05.20.Gg, 05.20.-y, 05.45.+b, 05.70.-a, 02.40.Vh, 02.40.-k}
Equilibrium properties
of an isolated Hamiltonian dynamical system with many, say $10^{24}$, 
degrees of freedom
is probably best described using a thermodynamic
formalism for a canonical ensemble at a
fixed temperature, even though this means
introducing fluctuations in an otherwise conserved quantity, the energy. 
For a more moderate number of degrees of freedom, say $10^3-10^8$,
numerical simulations become feasible,
and it is desirable to obtain a description 
in terms of the micro-canonical ensemble itself where the values
of the first integrals are fixed quantities.
In such an approach, 
geometrical properties of the level surfaces
reflect thermodynamic relations and,
by invoking the ergodic hypothesis, also dynamical properties
of the underlying system.
In particular, when energy is the only first integral,
measurements may then be done by time averaging 
(cf.\ \cite{Rugh97,Rugh98}).
The purpose of this article is to develop such a
micro-thermodynamic formalism
further, taking into account parameter dependency
and the presence of other first integrals. 
Within this framework we will also
(section \ref{sec adia})
discuss a natural formulation of a 1st law of thermodynamics,
or fundamental relation, based upon the bulk-entropy, $S_\Omega$.
We refer to
Abraham and Marsden \cite{AM}
as well as 
Landau and Lifshitz \cite {LL}
for a general
introduction to thermodynamic ensembles,
to Lebowitz et al. \cite{LPV} for an illustrative
example of some differences between the ensembles
and to Evans and Morriss \cite{EM} for
practical calculations carried out in the micro canonical ensemble.
We also refer to Jepps et al. \cite {JAE} where the presence of
other (approximative) first integrals is of relevance and
to Otter \cite {O} who studied reaction events 
using mixed ensemble averages.

\section{Micro-canonical ensembles}

 For simplicity, we consider
a Euclidean phase space,
 $\Omega \!=\! R^{2d}$, $d \geq 1$,
and a Hamiltonian function, $H : \Omega \rightarrow R$, 
bounded from below and of sufficient rapid growth at infinity. The dynamics
preserves the Liouville measure,
here the Lebesgue measure, $m\!=\!d^d\!x\, d^d\!p$.
There may be other first integrals,
denoted $F\!=\!F_1,\ldots,F_m$, $m \geq 0$ 
Note that in section \ref{sec deriv} we shall write
$F_0\!=\!H$ for the Hamiltonian which is then considered at the same footing
as the other first integrals. All first integrals are assumed
to be in involution.
We also assume that all functions are
known analytically and that the Hamiltonian
depends smoothly
on some external
real parameters, denoted $\Lambda\!=\!\Lambda_1,\ldots,\Lambda_n$. 
By contrast we do not allow the other first integrals to depend on $\Lambda$.
This is for technical reasons (cf.\ below), though in some cases
such a condition could be relaxed.
For fixed values of
 parameters, $\Lambda\!=\!\lambda$,
 of first integrals, $F\!=\!I$, and of the energy,
 $H_\lambda\!=\!E$,
the  subspace,
 $A\!=\!A[E,I,\lambda]\!=\!\{\xi\in R^{2d}:
  H_\lambda(\xi)\!=\!E, F(\xi)\!=\!I\}$,
 is invariant under the 
dynamics of $H_\lambda$. We will  assume that values 
are chosen so that
the differentials,
$dH$, $dF_1$,\ldots,$dF_m$,
are all independent on $A[E,I,\lambda]$. This  in particular implies that
$A$ is a smooth co-dimension $m+1$
sub-manifold of our Euclidean space.

In the literature one will find (at least) two 
definitions (denoted bulk and surface)
of a micro-canonical entropy and temperature.
It turns out that we shall need both.
 Thus we define
\begin{equation}
  e^{S_\Omega(E,I,\lambda)} \equiv \int m \; 
         \Theta \left(E\!-\!H_\lambda\right) \delta \left(
             I\!-\!F \right) \
  \label{eq:bulk}
\end{equation}
where $\Theta$ denotes the Heaviside function and
\begin{equation}
  e^{S_\mu(E,I,\lambda)} \equiv 
       \int m\; \delta \left( E\!-\!H_\lambda,I\!-\!F \right) .
  \label{eq:surface}
\end{equation}
The bulk- and the surface-temperature are then given by:
\begin{equation}
  \frac{1}{T_\Omega} = \frac{\partial S_\Omega}{\partial E} ,  \ \ \ \
  \frac{1}{T_\mu} = \frac{\partial S_\mu}{\partial E}  .
  \label{eq:temp}
\end{equation}
Derivatives with respect to other first integrals are
considered in Section (\ref{sec deriv}).
We also have generalized bulk- and surface-pressures: 
\begin{equation}
   p^i_\Omega={T_\Omega}
            \frac{\partial S_\Omega}{\partial \lambda_i},\ \ 
     \ \ 
   p^i_\mu={T_\mu}
        \frac{\partial S_\mu}{\partial \lambda_i},\ \ i=1,\ldots,n .
   \label{eq:conjugate}
\end{equation}

Taking an average
in the micro-canonical ensemble will here 
mean taking the surface-average, i.e.\
\begin{equation}
    \langle \phi \rangle_\mu \equiv
    \langle \phi | E,I,\lambda \rangle = 
       \frac{\int m\; \delta \left( E\!-\!H_\lambda,I\!-\!F \right) \phi}
            {\int m\; \delta \left( E\!-\!H_\lambda,I\!-\!F \right)}.
\end{equation}
In practice, $\langle \phi\rangle_\mu$, is often
calculated by time-averaging (assuming ergodicity), 
thus giving a
dynamical preference to the surface-average relative to other
ensemble-averages.

The calculation of either of the two entropies
may be difficult or even impossible when the number
of degrees of freedom in the system is large.
On the other hand, the associated  temperatures and generalized
pressures
may be probed using time-averaging (cf.\ Section \ref{sec deriv}).
A bulk-pressure,
$p^i_\Omega$,
 may be calculated as follows:
\begin{equation}
    p^i_\Omega =
       - \langle \frac{\partial H_\lambda}{\partial \lambda_i}
         \rangle_\mu =
       - \langle \frac{\partial H_\lambda}{\partial \lambda_i}
        | E,I,\lambda \rangle.
    \label{bulk-aver}
\end{equation}
To see this we note that the derivative of a Heavyside-function
yields a delta-function. It follows that
$1/T_\Omega=\partial S_\Omega / \partial E=e^{S_\mu}/e^{S_\Omega}$, and
therefore,

\begin{equation}
  p^i_\Omega={T_\Omega}
   \frac{\partial S_\Omega }{ \partial \lambda_i}=
 -  \frac{\int m\; \delta(E\!-\!H_\lambda,I\!-\!F)
       \frac{\partial H_\lambda}{\partial \lambda_i}}
  {\int m\; \delta(E\!-\!H_\lambda,I\!-\!F) }.
\end{equation}

By the very definition, there is always
a 1st law of thermodynamics for the bulk-entropy:
\begin{equation}
    {T_\Omega} dS_\Omega = dE + \sum_i p^i_\Omega \; d\lambda_i  .
    \label{dbulk}
\end{equation}
 The same kind of relation is, of course, valid for the surface-entropy but,
as we shall see, adiabatic invariance clearly  gives a preference to
the version involving the bulk-entropy:

\section{Adiabatic invariance}
\label{sec adia}
An adiabatic process is characterized by
a slow and smooth change in parameters
during which `thermal equilibrium' is maintained.
For example,  slowly moving a piston
of a cylinder containing a gas of particles,
In order to to avoid what is known as parametric resonance,
we fix a  smooth path in parameter
space, $s \in [0,1]
\mapsto c(s) \in R^n$,
 and traverse this path in rescaled time, i.e.\
$t \in [0,\Delta] \mapsto \lambda(t)=c(t/\Delta)$,
for some $\Delta>0$. A physical
trajectory, $\xi(t)$,
is evolved from an (almost) arbitrary point,
 $\xi(0)\in A[E,I,\lambda]$,
using the time- and $\Delta-$ dependent Hamiltonian,
$H_{\lambda(t)}$. The goal is to determine the micro-canonical state,
i.e.\ the values of $(E,I)$,
of the system at time $t=\Delta$.
The process is called adiabatic if this state has a well-defined
limit as $\Delta\rightarrow \infty$, the limit depending on the initial
state  $(E,I,\lambda)$ and the path, but not on the choice of initial point,
cf.\ Arnold \cite{Arnold},
section 52, or \cite{Simo}. 
Here again, we do not permit
the other first integrals to depend on $\Lambda$. 
Being time-independent and commuting with the Hamiltonian,
they therefore remain constant under the time-involution.
The energy, $E(t)=H_{\lambda(t)}(\xi(t))$,
being time-dependent
through $\lambda$, is in general  not constant in time. Instead we get
by Hamilton's equations:
$dE/dt=\sum_i \partial H_{\lambda_i} /\partial \lambda_i \cdot d\lambda_i/dt$,
and the `adiabatic' problem is then to see if $E(\Delta)$ has a well-defined
limit as $\Delta\rightarrow \infty$.

In order to attain the adiabatic limit,
$\Delta$ should be very large, in particular larger than the
time-scale, $\tau_{\rm erg}$, over which ergodic averaging takes place.
We may then look at the time-average of an observable
over an intermediate time-scale,
$\tau_{\rm erg} << \delta t << \Delta$.
This time-scale is short enough so that $\lambda$ does not change
significantly  but long enough so that a dynamical average may be replaced
by a micro canonical average. 
Under this `Adiabatic Ergodic Hypothesis' we get for the energy-change:
\begin{equation}
     \frac{1}{\delta t} \int_t^{t+\delta t} 
         \frac{dH_{\lambda(t)} \circ \xi(t)}{dt} dt
           \approx 
     \langle \frac{dH_{\lambda(t)}}{dt} | E(t), I,\lambda(t) \rangle .
\end{equation}

Over the time-scale, $\delta t$,
the $\lambda$-derivative is almost constant 
(because of the scaling with $\Delta$) and may therefore
be taken outside the average:
\begin{equation}
    \frac{E(t+\delta t)-E(t)}{\delta t} \approx 
     \sum_i \langle \frac{\partial H}{\partial \lambda_i} 
     | E(t),I,\lambda(t)\rangle
         \cdot \frac{d \lambda_i}{d t},
\end{equation}
an expression which is correct to order $\tau_{\rm erg}/\Delta$.
Taking the $\Delta\rightarrow\infty$ limit we get
an identity between differentials,
\begin{equation}
    dE = \sum_i \langle \frac{\partial H}{\partial \lambda_i}
               | E,I,\lambda \rangle \ d\lambda_i ,
\end{equation}
valid precisely in the adiabatic limit. 
By (\ref{bulk-aver}) and (\ref{dbulk}), 
we see that the bulk-entropy, $S_\Omega$, is
indeed an adiabatic invariant.
In 1+1 dimensions
the bulk-entropy is just the action integral and the phenomena is well-known
(adiabatic invariance of the action, cf.\ Arnold \cite{Arnold}).
In higher dimensions a similar result was obtained 
by Kazuga \cite{Kazuga}, though  in a different context.

 Now, the bulk-entropy is {\em a fortiori} strictly increasing
 as a function
of the energy (at given parameter- and first integral-values). 
One particular gratifying, though non-trivial (!), consequence is that
by traversing a loop  adiabatically
the energy 
must return to its original value.

It is easy to put 
the Adiabatic Ergodic Hypothesis into  a more rigorous form
(in terms of decay of correlation functions).
It is, however,  virtually impossible to  check analytically
if such a condition really holds in a given situation.
One serious problem is that
critical slowing down (meaning that $\tau_{\rm erg}$ diverges)
 occurs if one encounters an additional first
integral along the traversed path.
 On the other hand, singularities in the energy surface are
likely to pose problems only in low  dimensions (notably 1+1).

Should one allow other first integrals to depend on the
parameters, the bulk-entropy, as we have defined it, is in general no
longer an adiabatic invariant. 
We have not been successful in finding a good replacement
for the bulk-entropy (the reader is encouraged to try for himself)
and consequently not allowed such a parameter dependence.

\section{Surface - derivatives}
\label{sec deriv}
Close to equilibrium
we may express response functions as 
derivatives of averages with respect to parameters and
values of the first integrals 
(in this section this includes the Hamiltonian).
When the Hamiltonian is the only first integral and there 
is no parameter dependence,
it was shown in \cite{Rugh97,Rugh98} (cf.\  also \cite{JAE,RP}) that
energy-derivatives of a micro-canonical average
may themselves be calculated as averages.
The general case turns out to be quite similar.

Our assumption on the first integrals being independent means that
    in a neighborhood of
   the integral surface, $A[E,I,\lambda]$,
we may construct vector fields, $X_0$,\ldots,$X_m$,
   for which
\begin{equation}
    dF_i(X_j)=\delta_{ij}, \ \ \ i,j=0,\ldots,m .
	   \label{eq transver}
\end{equation}
Geometrically, the vector field, $X_i$, is transversal to the
$F_i=I_i$ surface but parallel to the other surfaces, $F_j=I_j$,
$j\neq i$.
We make the observation that if 
$f(x)$ is a (suitably smooth)
function and $V$ is a vector field for which
$df(V)\equiv 0$, then the Lie-derivative, $L_V \delta(f(x))=
 (V \cdot \nabla) \delta(f(x)) \equiv 0$,  vanishes identically.
This is clearly true if the delta-function were a smooth
function and the claim then follows by approximation
(a rigorous proof is quite lengthy).
This applies to our Hamiltonian set-up,
since for all
           $i,j=0,\ldots,m$
\begin{equation}
    (\frac{\partial}{\partial I_i} + (X_i \cdot \nabla))
           (I_j -F_j) = \delta_{ij} - \delta_{ij}= 0 .
\end{equation}
Hence, if $\phi$ is any smooth function,
\begin{equation}
  \int m\; \phi  \cdot
    (\frac{\partial}{\partial I_i} + (X_i \cdot \nabla))
  \delta(I\!-\!F)  = 0.
\end{equation}
Taking the $I_i$-derivative outside the integral and 
carrying the Lie-derivative out by partial integration we obtain

\begin{equation}
    \frac{\partial}{\partial I_i}
       \int m \ \delta(I-F)\ \phi =
     \int m \ \delta(I-F)\;   \nabla\! \cdot\! (\phi X_i) .
     \label{eq int deriv}
\end{equation}

The identity,
     $(\frac{\partial}{\partial \lambda_k}
     + \frac{\partial H_\lambda}{\partial \lambda_k} \cdot
     \frac{\partial}{\partial E }) \delta(I-F) \equiv 0
    $
(recall that only the Hamiltonian depends on $\Lambda$), also implies:
 \begin{equation}
     \frac{\partial}{\partial \lambda_k}\!\! 
            \int \!\! m\, \delta(I\!-\! F) \phi
       = 
     -\!\! \int \!\! m \, \delta(I\!-\! F) 
      \nabla\! \cdot\!
       (\phi  \frac{\partial H_\lambda}{\partial \lambda_k}\! X_0) .
       \label{eq ham deriv}
 \end{equation}

The surface entropy is the logarithm of a surface integral. 
Hence, when taking derivatives
a normalisation factor appears which precisely turns the derivative
into a micro-canonical average. Thus, for the 
 inverse `generalized' surface-temperatures,
 $\beta^i_\mu=
     \frac{\partial S_\mu}{\partial I_i}$ (with $\beta^0_\mu=1/T_\mu$), we get
  using (\ref{eq int deriv}) and setting $\phi=1$,
  \begin{equation}
     \beta^i_\mu = 
	     \frac
          {\frac{\partial}{\partial I_i}
              \int m \ \delta(I-F)\ \phi }
          {
              \int m \ \delta(I-F)\ \phi }=
             \langle \nabla \cdot X_i |E,I,\lambda \rangle ,
	     \label{eq gen temp}
  \end{equation}
Similarly for the `generalized' pressures, using (\ref{eq ham deriv}),
  \begin{equation}
     p^k_\mu = \frac{\partial S_\mu}{\partial \lambda_k}=
      -\langle \nabla \cdot (
      \frac{\partial H_\lambda}{\partial \lambda_k} \cdot
         X_0) |E,I,\lambda \rangle ,
	  \label{eq gen press}
  \end{equation}
When taking a derivative of an average 
and writing $\delta=\delta(I-F))$ we have
 \begin{equation}
    \frac{\partial}{\partial I_i}
          \langle \phi \rangle_\mu =
	  \frac
              {\frac{\partial}{\partial I_i} \int\! m \,\delta\, \phi}
	      {\int\! m \,\delta}
	  -\frac
              {\frac{\partial}{\partial I_i} \int\! m \,\delta\, \phi}
	      {\int\! m \,\delta} \;
	  \frac
              {\frac{\partial}{\partial I_i} \int\! m \,\delta }
	      {\int\! m\, \delta}
  \end{equation}
 which by the definition of averages and generalized temperatures reduces to
 \begin{equation}
    \frac{\partial}{\partial I_i}
          \langle \phi \rangle_\mu =
          \langle \nabla \cdot (\phi X_i) \rangle_\mu - \beta^i_\mu
                     \langle \phi \rangle_\mu  s.
  \end{equation}
We calculate in the same way,
 \begin{equation}
     \frac{\partial}{\partial \lambda_k} 
          \langle \phi \rangle_\mu =
        -  \langle \nabla \cdot (\phi 
        \frac{\partial H_\lambda}{\partial \lambda_k} X_0) 
                         \rangle_\mu - p^k_\mu
                     \langle \phi \rangle_\mu  .
      \label{eq gen deriv}
 \end{equation}
These thermodynamic identities provide the natural generalizations
of the results found in \cite{Rugh97,Rugh98}.
In the above formulae 
the other first integrals may, in fact,
be allowed to be parameter dependent 
(essentially because there is no Heavyside function in the above).
The straight-forward
derivation of formulae in this case is no more difficult
and left to the reader.

We also note that if the Liouville measure has the form
$dm = \rho (\xi) d^{2d}\xi$ (for instance in local coordinates 
on a symplectic manifold), the only change in the above
formulae is to replace the divergence of a vector field,
$\nabla \cdot V$, by
$\frac{1}{\rho} \nabla \cdot (\rho V)$ (cf.\  \cite{Rugh98}).

\section{Bulk - derivatives}
\label{sec bulk deriv}
Should one wish to take
derivatives in the bulk-ensemble 
the procedure is slightly different.
Restricting our attention to the energy derivative,
suppose that 
$Y$ is a smooth vector field such that [compare with (\ref{eq transver})]:
 \begin{equation}
    \nabla\! \cdot\! Y \equiv 1 \ \ \ {\rm and} \ \ \
     dF_i(Y) \equiv 0, \ 1\leq i\leq m.
      \label{eq bulk trans}
   \end{equation}
One may certainly find such $Y$ when 
there are no other first integrals present. In the general
case it is less clear because here we need 
the vector field to be defined `smoothly' throughout the `bulk'.
 Assuming, however,  that we have found such a vector
field  we note that
\begin{equation}
   e^{S_\Omega} = \int m\; (\nabla \cdot Y)\; \Theta(E-H)\delta(I-F)
\end{equation}
which by partial integration yields
\begin{equation}
   e^{S_\Omega} = \int m \;dH_\lambda(Y) \;\delta(E-H,I-F)
\end{equation}
In particular, we obtain the 
following (well-known, when energy is the only first integral)
 formula for the bulk-temperature
\begin{equation}
   T_\Omega = e^{S_\Omega-S_\mu} = 
       \langle dH_\lambda(Y) | E,I,\lambda \rangle.
        \label{eq bulk temp}
\end{equation}
For comparison we note that in
the canonical ensemble
where an integral of $\phi$ looks as follows:
\begin{equation}
   \int m \; e^{-\beta H_\lambda} \; \delta(I-F) \; \phi,
\end{equation}
the same calculation shows that
\begin{equation}
    T_{\rm can}= \beta^{-1} = 
       \langle dH_\lambda(Y) | \beta,I,\lambda \rangle.
\end{equation}
In the literature 
$dH(Y)$ is therefore often used to `define'   the
micro-canonical temperature (again measured by time-averaging).
With no other first integrals present one may take $Y$ to be
the canonical momenta divided by the number of degrees of freedom, i.e.\
$Y=P/d$.  In that case, 
$dH(Y)=\frac{1}{d} dH(P)= 1/d \sum \dot{q_i} p_i$ is the
normalised reduced action (proportional to the kinetic energy, when 
quadratic in momenta). By time-averaging, we obtain a quantity proportional
 to the reduced action-integral. 
This is extremal under variations of the
trajectory preserving the energy
(Maupertius principle, \cite[Section 45]{Arnold}). 
One would thus expect a finite  time-average of $dH(P)$ to
`probe' a larger neighborhood of
the trajectory than time-averages of other observables.
This suggests a faster ergodic averaging, whence
stronger numerical stability, when calculating time-averages of $dH(P)$
relative to other observables of the same ensemble-variance.

\section{An example}
Consider an ensemble of $N$ particles
 moving on a 
3-dimensional 
torus, $(R/Z)^3$,
under the influence of pair-potentials.
Putting things on a torus compactifies the configurational space
but otherwise does not affect our results.
The Hamiltonian,
$H=\sum_{i=0}^N {\bf p}_i^2/2m_i + \sum_{i,j}U({\bf x}_i-{\bf x}_j)$,
is translational invariant, hence the total momenta,
${\bf p}_{\rm tot}=\sum {\bf p}_i$, provides $3$ first integrals in 
addition to the
Hamiltonian. If one here uses the normalised  total kinetic energy 
as a measure of the temperature one runs into the following paradox:
A configuration which
 minimizes the
potential energy but has all particles
moving at the same constant velocity, ${\bf v}_0$,
is stationary. 
With no `apparent activity' one should assign a temperature zero
to this configuration but this is clearly not 
what the total kinetic energy does.

One solution to this paradox suggests itself:
Move to a frame where the average velocity vanishes.
Such a procedure is, in fact, a natural consequence of
our micro-canonical formalism. To get the pre-factors right
we carry out the details:
If $P=({\bf p}_1,\ldots,{\bf p}_N,{\bf 0},\ldots,{\bf 0})$ denotes the 
full momentum vector in $R^{3N} \times(R/Z)^{3N}$ 
and ${\bf v}=\sum_i {\bf p}_i/\sum_i m_i$ is
the average velocity then the reduced momentum vector and kinetic energy
is given by
$Z=({\bf p}_1-m_1 {\bf v},\ldots,{\bf p}_N-m_N {\bf v},
 {\bf 0},\ldots,{\bf 0})$ and 
$K_{\rm red} = \sum_i ({\bf p}_i-m_i {\bf v})^2/2m_i$, respectively.
Straight-forward calculations give $dH(Z)=2K_{\rm red}$ and
$d{\bf p}_{\rm tot}(Z)\equiv 0$.
The latter is
precisely the requirement of `parallelism' needed for both (\ref{eq transver})
and (\ref{eq bulk trans}).
Furthermore, we have
$\nabla\! \cdot\! Z=3(N\!-\!1)$,
and 
$\nabla\! \cdot\! (Z/dH(Z))=(3(N\!-\!1)\!-\!2)/(2 K_{\rm red})$. 
After the appropriate normalization,
formulae (\ref{eq gen temp}) and (\ref{eq bulk temp}) yield the surface- and
bulk-temperatures:
  \begin{equation}
     \frac{1}{T_\mu}=\langle 
         \frac{3(N\!-\!1)\!-\!2}{2 K_{\rm red}}\rangle_\mu\ \ \
     \mbox{and} \ \ \
     T_\Omega = 
     \langle \frac{2 K_{\rm red}}{3(N\!-\!1)}\rangle_\mu.
   \end{equation}
In both cases the conservation of 3 momenta results in
a subtraction of 3 degrees of freedom as compared to the
unconstrained case (pleasing on physical grounds).
It is interesting also to note that the above
formulae involving the reduced kinetic energy are valid irrespective
of the total momentum being preserved or not ! When the total momentum
is only approximately preserved one could use the above construction to
define an approximate `instantaneous' and possibly `local' temperature.
In \cite{JAE}, such a problem was considered and the solution
suggested was to choose
a vector field, $Z$, depending on the configurational coordinates only.
This automatically ensures `parallelism', i.e.\
$d{\bf p}_{\rm tot}(Z)\equiv 0$,
and leads to what is often denoted the `configurational temperature'.
On the other hand, for numerical reasons the reduced kinetic energy
may be preferable.

The case of a conserved angular momentum  (an experiment involving
an axial symmetry) follows the same
straight-forward procedure and is left to the reader.

\end{document}